\documentclass[useAMS,usenatbib]{mn2e}

\usepackage{graphicx}
\usepackage{gensymb}
\usepackage{natbib}
\usepackage{amssymb}
\usepackage{upgreek}
\usepackage{array}
\usepackage{multirow}
\usepackage[usenames]{xcolor}

\title[The magnetic inclination of $\gamma$-ray-loud pulsars]{Investigating the magnetic inclination angle distribution of $\gamma$-ray-loud radio pulsars}
\author[S. C. Rookyard, P. Weltevrede and S. Johnston]{S. C. Rookyard$^{1}$\thanks{E-mail:
simon.rookyard@postgrad.manchester.ac.uk}, P. Weltevrede$^{1}$ and S. Johnston$^{2}$\\
$^{1}$Jodrell Bank Centre for Astrophysics, School of Physics and Astronomy, University of Manchester, Manchester M13 9PL, UK\\
$^{2}$CSIRO Astronomy and Space Science, Australia Telescope National Facility, Epping NSW 1710, Australia}
\begin{document}

\date{Accepted 1988 December 15. Received 1988 December 14; in original form 1988 October 11}

\pagerange{\pageref{firstpage}--\pageref{lastpage}} \pubyear{2002}

\maketitle

\label{firstpage}

\begin{abstract}
Several studies have shown the distribution of pulsars' 
magnetic inclination angles to be skewed towards low values compared 
with the distribution expected if the rotation and magnetic axes are placed randomly on the star. Here we focus on a sample of 
28 $\gamma$-ray-detected pulsars using data taken as part of 
the Parkes telescope's \emph{FERMI} timing program. In doing 
so we find a preference in the sample for low magnetic inclination 
angles, $\alpha$, in stark contrast to both the expectation that 
the magnetic and rotation axes are orientated randomly at the 
birth of the pulsar and to $\gamma$-ray-emission-model-based 
expected biases. In this paper, after exploring potential 
explanations, we conclude that there are two possible causes 
of this preference, namely that low $\alpha$ values are intrinsic 
to the sample, or that the emission regions extend outside what 
is traditionally thought to be the open-field-line region in 
a way which is dependent on the magnetic inclination. Each possibility 
is expected to have important consequences, ranging 
from supernova physics to population studies of pulsars and considerations 
of the radio beaming fraction. We also present a 
simple conversion scheme between the observed and intrinsic magnetic 
inclinations which is valid under the assumption that the observed 
skew is not intrinsic and which can be applied to all existing measurements.
We argue that extending the active field-line region 
will help to resolve the existing tension between emission 
geometries derived from radio polarisation measurements and those 
required to model $\gamma$-ray light curves. 
\end{abstract}


\section{Introduction}
\label{SectIntroduction}

The recent increase in the number of $\gamma$-ray pulsar detections, largely due to the \emph{FERMI} satellite, allows great progress to be made in our understanding of these sources. A key question is that of the location and structure of the emission region. An important element required to test various models of this against observations is the determination of the ``viewing geometry'', which describes how an observer's line of sight samples a given pulsar's magnetosphere. Knowledge of this for a particular pulsar allows predictions to be made corresponding to a given model of the $\gamma$-ray beam pattern, which can then be compared to observations in order to examine the model's veracity. Based on radio polarisation data, viewing geometry constraints were presented by \cite{rwj14a} (henceforth RWJ14) for 28 pulsars which were included in the 2nd \emph{FERMI} Large Area Telescope catalogue of $\gamma$-ray pulsars \citep{aaa+13}. To this end, data taken predominantly at 1369~MHz using the Parkes radio telescope as part of the \emph{FERMI} timing project \citep{wjm+10} were obtained. The aim of this timing project is to make regular monthly radio observations of a set of pulsars with a high energy loss rate $\dot{E}$ to construct timing models allowing to tag rotational phases to photons detected by the \emph{FERMI} satellite. In RWJ14 these data were used to construct polarisation-calibrated averaged pulse profiles of the total, linearly polarised and circularly polarised intensity, along with profiles of the position angle (PA) of the linearly polarised component of the observed emission.

The viewing geometry is characterised by two angles, the magnetic inclination $\alpha$ (the angle between the rotation and magnetic axes) and the impact parameter $\beta$ (the angle between the line of sight and the magnetic axis at the closest approach). These parameters can be constrained by analysing the polarised radio emission and the pulse profile of the pulsar.

\begin{figure}
\centering
\includegraphics[height=\hsize,angle=270]{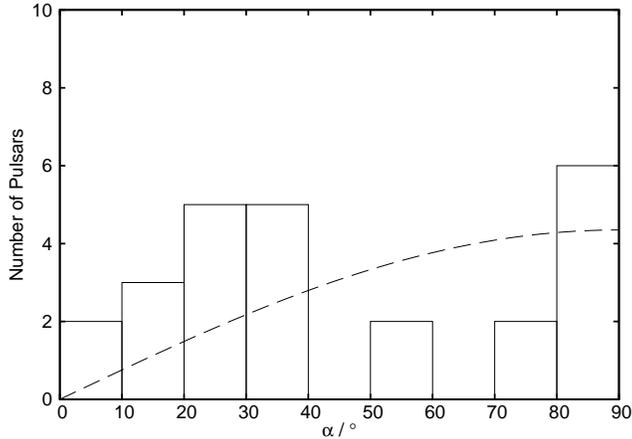}
\caption{\label{FigOriginalDistribution} Distribution of most likely $\alpha$ values, as quoted in Table 2 of RWJ14 and shown in Fig.~29 of the same paper. The dashed curve is the sinusoidal distribution expected for a randomly orientated rotation axis and line of sight in the absence of any observational biases. It can clearly be seen that the distributions are different. }
\end{figure}

Fig.~\ref{FigOriginalDistribution} shows the distribution of $\alpha$ values presented in Table 2 of RWJ14. For this figure and subsequent analysis $\alpha$ was mapped between 0 and $90\degree$ ($\alpha \rightarrow 180\degree - \alpha$ if $\alpha > 90\degree$) as there is no physical difference between these two regimes. For three out of 28 sources the viewing geometry could not be reliably determined due to extreme uncertainties on the relevant parameters, as discussed in $\S$~4 of RWJ14. 

This distribution shows a significant tendency towards low 
values such that 60\% of the sample have $\alpha < 40\degree$, consistent 
with other results in the literature. \cite{tm98} discussed 
$\alpha$ distributions using values derived by \cite{gou94} 
and \cite{ran90}, all of which demonstrate a peak at 
$\alpha \sim 40\degree$. \cite{tm98} also separated 
the sample according to characteristic age ($\tau_c = P/2\dot{P}$, which is not necessarily the true age) and corrected the measured distributions according to the beaming fraction, 
the proportion of the celestial sphere covered by the emission beam, 
assuming an empirical relationship between $\rho$ and $P$. 
One of these distributions comprised pulsars younger than $10^{6.5}$ yr, which is approximately the same as the range of ages of our sample. 
This inferred intrinsic distribution also exhibited a tendency towards low 
$\alpha$ values, consistent with our results. The methods used by RWJ14 were 
different from those used by \cite{tm98}, supporting the determined shape 
of the observed $\alpha$ distribution.

This is a puzzling result, however. It is commonly assumed that 
the relative orientation of the magnetic and rotation axes is random at birth, in the sense that the two are placed randomly on the star. If 
this is the case, the intrinsic $\alpha$ distribution should be 
sinusoidal (e.g., \citealt{gh96}), as shown by
the dashed curve in Fig.~\ref{FigOriginalDistribution}.  
Two effects suggest that any deviation from the 
sinusoidal distribution should in fact be a bias towards high $\alpha$ 
values, opposite to the apparent bias in this sample. The beaming 
fraction, which is equivalent to the probability that the beam of a given pulsar intersects 
an observer's line of sight, increases with $\alpha$ (see e.g. \citealt{tm98}). This means that radio observations should detect a 
larger proportion of the total population of orthogonal rotators 
than of aligned rotators. Additionally, the prediction from $\gamma$-ray 
models is that the pulsations of $\gamma$-ray-loud pulsars are 
more easily detected when $\alpha$ is large (e.g., \citealt{rw10, wrw+09}). 
As the pulsars in this sample are all $\gamma$-ray-detected, we 
would expect this selection effect to skew our observed distribution 
further towards higher values, by an amount which depends on the 
particular $\gamma$-ray model. Neither this nor the effect of 
the beaming fraction will be quantified further in this paper, 
but their effect on our results will be discussed.   

An immediate question raised by the discrepancy between the observed
distribution and our expectation is whether the $\alpha$ values should
be expected to follow the birth distribution. It has been suggested that 
the magnetic axis should align with the rotation axis over time, meaning
this assumption will only be valid for a sample of sufficiently young 
pulsars. However, \cite{wj08a} showed that for the pulsar population 
as a whole, if the axes are assumed to be randomly aligned at birth, 
the proportion of pulsars exhibiting an interpulse as a function of 
age is best reproduced by allowing the angle between them to decrease 
with a timescale $\sim 7 \times 10^7$~yr. Other estimates of this timescale 
are $\sim 10^6$ to $\sim 10^7$~yr \citep{ycb+10, tm98}. The highest 
characteristic age of any pulsar in this sample is $10^{5.7}$~yr 
(PSR J1057--5226), which is significantly less than the alignment 
timescale. This indicates that the distribution for this sample 
should not have evolved significantly from the birth distribution. 
Further to this, it has recently been suggested \citep{lgw+13} that 
the magnetic axis of the Crab pulsar ($\tau_c\simeq$~$10^{3.1}$~yr) may be 
moving \emph{away} from the rotation axis. If all very young pulsars 
undergo such a period of increasing $\alpha$, this would strengthen 
the expectation that large $\alpha$ values should be favoured for 
the sample considered here.

In this paper we discuss various observational and other biases which could potentially cause the observed tendency towards low $\alpha$ values, including emission generated outside what is traditionally thought to be the open-field-line region, 
and find two possible explanations for the form the distribution takes. The first of these 
is simply that the magnetic and rotation axes are not randomly orientated 
at birth as expected, but instead neutron stars are more likely to be born with the axes aligned than orthogonal. Alternatively, 
the axes \emph{are} randomly orientated and one of the assumptions used in the derivation of the $\alpha$ values is invalid. We argue that the most likely possibility is that the 
radio emission beam is larger than expected for a dipolar field 
by a proportion which is dependent on the magnetic inclination.

\section{Method to constrain the viewing geometry}
\label{SectTheory}

In the following text we give a summary of the process for constraining the viewing geometry, as applied in RWJ14 and other publications in which radio polarisation data are interpreted. This will form the mathematical basis for the discussion in this paper.  

Information about the viewing geometry can be obtained by fitting the Rotating Vector Model (RVM; \citealt{rc69}) to the observed PA curve of each pulsar. The RVM depends on $\alpha$ and $\beta$ and so the resulting $\chi^2$ surface in ($\alpha$, $\beta$) space constitutes an initial constraint on these parameters.

The viewing geometry can be constrained further by considering the effect of aberration and retardation (A/R; \citealt{bcw91, drh04}) and the emission height ($h_{\mathrm{em}}$, the distance of the emission region from the centre of the star) which can subsequently be determined. This effect arises from relativistic motion of the emission region relative to the observer as the region corotates with the neutron star. The PA curve predicted by the RVM features a point of inflection which, neglecting the A/R effect, would be observed when the fiducial plane (the plane containing the two axes) passes the line of sight. However, the A/R effect results in a delay in pulse phase of the inflection point relative to the location of the fiducial plane inferred from the intensity profile of   

\begin{equation}
\label{EqOffset} \Delta\phi = \frac{8\pi h_{\mathrm{em}}}{Pc}   ,
\end{equation}
where $P$ is the rotation period of the star and $c$ is the speed of light \citep{bcw91}. The pulse phase at which the inflection point of the observed PA curve occurs follows from RVM fitting, and the location of the fiducial plane in terms of pulse phase can be judged somewhat subjectively using the profile morphology. The relative delay $\Delta\phi$ then follows and so $h_{\mathrm{em}}$ can be determined.  

With the emission height known, the half-opening-angle, $\rho$, of the beam of radio emission can be found by assuming that the beam is bounded by tangents to the last-open-field lines of a dipolar magnetic field. Given this assumption, $h_{\mathrm{em}}$ yields the half-opening angle via    

\begin{equation}
\label{EqBeamHalfOpeningAngle} \rho = \theta_{\mathrm{PC}} + \arctan\left(\frac{1}{2} \tan\theta_{\mathrm{PC}}\right)   ,
\end{equation}
where $\theta_{\mathrm{PC}}$, the angular radius of the open-field-line region, 
is given by   
\begin{equation}
\label{EqThetaPC} \theta_{\mathrm{PC}} = \arcsin\left(\sqrt{\frac{2\pi h_{\mathrm{em}}}{Pc}}\right)   
\end{equation}
(e.g., \citealt{PulsarAstronomy}). This beamwidth can finally be related to the range of rotational phase for which the line of sight samples the open-field-line region, $W_{\mathrm{open}}$, and the viewing geometry by   

\begin{equation}
\label{EqRhoContours} \cos\rho = \cos\alpha\cos(\alpha + \beta) + \sin\alpha\sin(\alpha + \beta)\cos\left(\frac{W_{\mathrm{open}}}{2}\right)   
\end{equation}
\citep{ggr84}. The symmetry of the open-field-line region of a dipolar field means that the line of sight will sample open field lines for an equal amount of phase before and after the fiducial plane. In RWJ14 the fiducial plane was not necessarily chosen to correspond to the centre of the observed pulse, meaning that $W_{\mathrm{open}}$ cannot be assumed to be the same as the observed pulse width. Instead, it was taken to be twice the difference in phase between the fiducial plane position and the pulse edge furthest from it. The implications of this decision will be discussed in $\S$~\ref{SectMovingPhiFid}. The pulse edges were in general taken to be the points at which the emission was 10\% of the peak intensity. From the calculated values of $\rho$ and $W_{\mathrm{open}}$, Eq.~\ref{EqRhoContours} can be used to constrain $\alpha$ and $\beta$. The best joint solution from RVM fitting and the pulse width considerations was defined as the favoured viewing geometry.   

There are several possible reasons why an intrinsically 
sinusoidal distribution may appear skewed towards low 
$\alpha$ values. These include 
\begin{enumerate}
  \item a systematic underestimation of the phase of the PA curve inflection point;
  \item a systematic overestimation of the phase of the fiducial plane;
  \item a non-circularity of the emission region, and
  \item extra-cap emission (in which the emission region is larger than expected for the open-field-line region of a simple dipolar magnetic field).
\end{enumerate}

In the following subsections we consider each effect 
in turn and investigate whether 
it could sufficiently distort the $\alpha$ distribution. We will argue 
that the most plausible solution which allows the $\alpha$ distribution 
to be sinusoidal involves extra-cap emission for which the ratio of 
the beam radius to polar cap radius is $\alpha$-dependent.

\section{Considerations of possible biases}

\subsection{Systematic underestimation of the inflection point position}
\label{SectMovingPhi0}

\begin{figure}
\centering
\includegraphics[height=\hsize,angle=270]{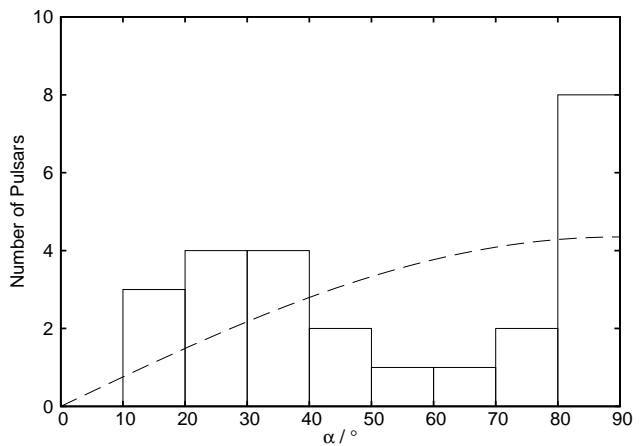}
\caption{\label{FigShiftinf+2.2} Distribution of $\alpha$ values with an artificial offset of $+$2.2$\degree$ applied to the inflection point of each pulsar. The curve is the sinusoidal distribution which we (unsuccessfully) tried to reproduce. }
\end{figure}

The first effect we considered was a scenario in which the inflection point of the PA curve was 
systematically determined to be at an earlier pulse phase than its 
intrinsic location. Such a bias in the positioning of the inflection 
point could potentially be caused by scattering, which smears out the PA curve
\citep{k09, kj08}. Another potential cause of a bias could be noise, which will be discussed later in this subsection. A systematic underestimation of the location of the inflection point would mean that the determined 
offset between the fiducial plane and the inflection point would 
be underestimated. Then, from Eqs.~\ref{EqBeamHalfOpeningAngle} and \ref{EqThetaPC}, 
the emission height and as a consequence the predicted beamwidth would be 
underestimated, leading to an underestimation of $\alpha$ 
(Eq.~\ref{EqRhoContours}). 

To investigate the effect on the $\alpha$ distribution the inflection 
point of each of the 25 pulsars which contributed to the measured 
$\alpha$ distribution was offset to later phase by a set 
number of degrees. The fiducial plane and pulse edges were kept the same. 
The Kolmogorov-Smirnov (KS) test, which is independent of binning, was used to quantify the difference between the resulting $\alpha$ distribution and the sinusoidal distribution. A lower test result indicates a greater likelihood that the two sets of values are drawn from different parent distributions. 

Fig.~\ref{FigShiftinf+2.2} shows the distribution obtained when an offset 
of $+2.2\degree$ was applied. This was the magnitude of offset for which 
the distribution was found to most closely resemble a sinusoidal 
distribution, with a KS test result of 4.6\% corresponding 
to a significance slightly above 2$\sigma$. Such an offset 
is similar in magnitude to the effect of 
scattering observed by \cite{kj08} in the case of PSR J0908--4913, 
suggesting the offset applied in the figure could potentially be 
explained by the effect of scattering in our sample. 
It can be seen that the peak at low $\alpha$ persists, 
accompanied by an excess of values relative to the sinusoidal distribution 
at $\alpha > 80\degree$. The distribution was qualitatively similar even 
when offsets were considered which were too large to be explained by 
the amount of scattering exhibited in the profiles. Although the $\sim$~2$\sigma$ significance shows that the distribution is formally consistent with a sinusoidal distribution, it is somewhat marginal given that the match is not very good even after optimising the shift, which was a free parameter. 
Furthermore, the $\gamma$-ray selection effect and radio beaming fraction biases in our sample (see $\S$~\ref{SectIntroduction}) are such that the observed 
distribution should be skewed towards high-$\alpha$ relative to the sinusoidal 
distribution. This cannot be reproduced by the shift. Hence 
we argue that, whilst scattering may play a role, it is not the main 
cause of the difference.

The effect of S/N on the determination of the inflection point was 
also investigated. The inflection point is typically at later phase than the midpoint of the profile, meaning there are more significant PA values before the inflection point. As noise contributes to the measured PA this asymmetry might, in principle, result in a bias in the determined inflection point. Polarised profiles were simulated according to the RVM using various combinations of S/N and offset between the inflection point and fiducial plane. White noise was added and the RVM was fitted to the resulting PA curve. In all cases the inflection point position was unaffected when averaged over a large number of simulations, indicating that noise does not cause a systematic bias. It is therefore apparent that if the intrinsic $\alpha$ 
distribution is sinusoidal the observed distribution cannot be explained 
by either interstellar scattering or the effects of noise on the PA swing.

\subsection{Systematic overestimation of the fiducial plane}
\label{SectMovingPhiFid}

\begin{figure}
\centering
\includegraphics[height=\hsize,angle=270]{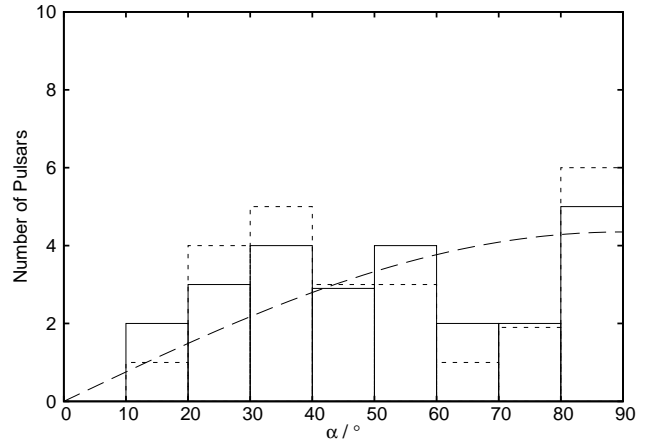}
\caption{\label{FigShiftfid-14} Distribution of $\alpha$ values with an artificial shift of $-14\degree$ applied to the location of the fiducial plane (solid line) and with the fiducial plane placed 12\% of the observed pulse width before the pulse for each pulsar (dotted line). The distributions are similar, with the former being a slightly closer match to the sinusoidal distribution (dashed curve). }
\end{figure}

Another potential cause of an excess of low measured $\alpha$ values is a 
systematic overestimation of the phase at which the fiducial plane 
is located in the profile. In $\S$~4 of RWJ14 the consequences of misplacing the fiducial plane were considered, but for single-component pulsars only; in this subsection we apply the same effect to the whole of the sample. It is important to consider this possibility as the fiducial plane of each pulsar was estimated subjectively, based on the profile shape. 

There are several possible causes of such a systematic overestimation. Within the context of the core-cone model \citep{ran93}, such an effect could arise if, for example, 
the leading portion of the beam is less intense than the trailing portion, such that one or multiple leading components are unobserved. This could be an extremely pronounced version of the preferential illumination at later phases noted for young pulsars by 
\cite{jw06} and discussed in the context of cyclotron absorption by \cite{flm03}. Alternatively the radio emission might not be generated in the ordered structure postulated by the core-cone model, but instead be generated in `patches' distributed randomly across the open-field-line region \citep{lm88}. 
It should be noted that a random distribution of patches does not explain the 
observed excess of low $\alpha$ values as a systematic bias is required (i.e., preferential illumination of the trailing side of the beam). 

Another possible reason for an overestimation of the fiducial plane position would be that the centre of the radio beam (in either the core-cone or patchy beam model) trails the magnetic axis (that is, the fiducial plane is not at the centre of the range of phase described by $W_{\mathrm{open}}$). However, in this situation only the relative offset of the fiducial plane and inflection point would be affected (the extent of the open-field-line region, and hence $W_{\mathrm{open}}$, would be unaffected), making this possibility equivalent to the inflection point offsets considered in $\S$~\ref{SectMovingPhi0}.

The result of a systematic overestimation of the fiducial plane was first of all investigated by offsetting the fiducial 
plane determined in RWJ14 by a set amount of pulse phase and determining the resulting $\alpha$ distribution, 
using a method analogous to that described in $\S$~\ref{SectIntroduction}. 
The distributions obtained using various offsets showed that as 
the offset becomes larger the increasing beamwidths cause 
the peak of the distribution to move towards higher $\alpha$. 
However, as the magnitude of the offset is increased further, 
the effect of the increasing inferred $W_{\mathrm{open}}$ values 
becomes dominant and the peak of the distribution moves towards 
lower $\alpha$.

Fig.~\ref{FigShiftfid-14} shows the $\alpha$ distribution (solid bars) obtained 
with an offset of --14$\degree$ applied to each pulsar. The data visibly match the sinusoidal distribution well, which was confirmed by a KS test result of 52\%. This implies that if for every pulsar in our sample the fiducial plane is 14$\degree$ earlier than thought based on the profile morphology, this might explain the shape of the observed $\alpha$ distribution.

One might na\"{i}vely expect an overestimation of the fiducial plane position to be related to the pulse width. A similar investigation to that detailed above was performed, in which the fiducial plane was placed before or after the start of the observed pulse by the same proportion of the pulse width. This corresponds to a scenario whereby the same proportion of the beam is illuminated for each pulsar. However, this did not result in any better match to a sinusoidal distribution. The results of the KS test peaked at 26\% for a fiducial plane position 12\% of the observed width before the start of the pulse, equivalent to the trailing 45\% of each pulsar's beam being illuminated. This is shown by the dotted bars in Fig.~\ref{FigShiftfid-14}.

The KS test value is lower for a fractional offset of the fiducial plane than for an absolute-offset, indicating a slight (but not significant) preference for the latter scenario. It therefore appears that our sample could have an 
intrinsically sinusoidal $\alpha$ distribution, provided either that the fiducial plane position is habitually $\sim 14\degree$ earlier than is suggested by the shape of the profile or that slightly less than the trailing half of the open-field-line region is illuminated. An important point to note is that neither of these scenarios is able to produce distributions with substantial excesses relative to the sinusoidal distribution at $\alpha > 80\degree$. As a result, accounting for the expected selection effects of $\gamma$-ray detectability and radio beaming fraction (see $\S$~\ref{SectIntroduction}) would be difficult. The plausibility of a fiducial plane offset will be discussed further in $\S$~\ref{SectImplicationsPhiFidOffset}.

\subsection{Emission height gradient}

\cite{gg01} proposed that, at a given frequency, the emission height of radiation could be greater further from the magnetic axis, and hence further from the centre of the profile. This means that the A/R effect will be most pronounced at the edges of the profile. It can be seen from Eqs.~\ref{EqBeamHalfOpeningAngle} and \ref{EqThetaPC} that it is the emission height at the edge of the beam which leads to the overall half-opening angle of the beam. In RWJ14, however, the fiducial plane position was estimated according to the locations of component peaks, which are not at the edge of the beam and hence, according to \cite{gg01}, will have a lesser emission height. This will have caused an underestimation of $\Delta\phi$ and therefore a systematic bias towards late fiducial plane positions. It can be seen from Fig.~4 of \cite{gg01} that for PSR~B0329+54 (which has a period of 0.7~s) this effect is $\sim$~4$\degree$. Considering Eq.~\ref{EqOffset} and noting that almost all the pulsars in our sample have periods less than 0.2~s, we could expect the underestimation of the fiducial plane to be greater for our sample. In addition, it is possible that high-$\dot{E}$ pulsars emit over a more extended altitude range \citep{wj08a}, potentially increasing this effect. 

The variation of the A/R effect across the profile will also move the ends of the PA curve towards later phase compared with the central region, which will distort the PA curve such that the leading half is made steeper while the trailing half is made more shallow. If not corrected for, this will cause a slight underestimation of the inflection point phase. However, as it is the ends of the PA curve at which the distortion is most pronounced, the curve will be less distorted close to the inflection point and so the observed position of the steepest gradient will change by only a small amount. The magnitude of this effect on $\phi_0$ varies between pulsars, but for this sample is expected to be typically $\sim$~1$\degree$. Therefore, the effect on our fiducial plane estimate will dominate over this effect and any change to the inflection point position can be neglected. 

An emission height gradient has the potential to make our observations consistent with a sinusoidal distribution. We can see from Eq.~\ref{EqOffset} that the --14$\degree$ offset in the fiducial plane position suggested in $\S$~\ref{SectMovingPhiFid} could be caused by a 0.06$R_{\mathrm{LC}}$ difference in the emission height, where $R_{\mathrm{LC}}$ is the light cylinder radius. In other words, the points in the profiles which were used to determine the favoured position of the fiducial plane should have been emitted 6\% of the light cylinder radius lower than were the edges of the pulse.

If an emission height gradient is the main cause of the observed tendency towards low $\alpha$ values, then choosing the fiducial plane position based on the pulse edges should lead to an $\alpha$ distribution which more closely resembles a sinusoid than Fig.~\ref{FigOriginalDistribution}. To test this, we obtained $\alpha$ values by positioning the fiducial plane at the midpoint of the observed pulse. For this we excluded pulsars for which we believed a significant fraction of the beam is not illuminated\footnote{The pulsars used were PSRs J0631+1036, J0659+1414, J0729--1448, the interpulse of J0908--4913, J0940--5428, J1105--6107, J1112--6103, J1119--6127 (case (a)), J1420--6048, J1531--5610, J1648--4611, J1702--4128, J1709--4429 and J1718--3825. See RWJ14 for motivation of these choices.}. 

Despite its promise, the results of this test showed little change in the $\alpha$ values, with marginally more pulsars decreasing in $\alpha$ than increasing (as described in $\S$~\ref{SectMovingPhiFid}, there are two competing effects acting on the $\alpha$ values). This indicates that, whilst we do not rule out an emission height gradient for these pulsars, such a gradient cannot explain the excess of low $\alpha$ values. This means that the potential --14$\degree$ offset determined in the previous subsection would require some other physical justification.

\subsection{Non-circularity of the emission region}
\label{SectNonCirc}

Another possible cause of a systematic underestimation of $\alpha$ is the assumption that the beam of emission is confined to a circular open-field-line region. To investigate the effect of other beam shapes, a model was considered in which the emission region was elongated into an ellipse. The methodology to determine $\alpha$ for elliptical beams is analogous to the case of circular beams, except that the relevant half-opening angle in Eq.~\ref{EqRhoContours} is $\rho_{\mathrm{ell}}$, which is a function of the magnetic longitude at which the line of sight enters and exits the beam. We define the beam such that the half-opening angle described by tangents to the last open field lines, $\rho$ (Eq.~\ref{EqBeamHalfOpeningAngle}), forms the semi-minor axis of this ellipse. This makes $\rho_{\mathrm{ell}}$ larger than the prediction from Eq.~\ref{EqBeamHalfOpeningAngle}, regardless of the direction in which the beam is elongated. This systematic underestimation of the beam half opening angle as derived under the assumption of a circular beam causes a systematic underestimation of $\alpha$.

Two variants of this elliptical beam model were examined. In the first 
of these the major axis of the beam lay in the fiducial plane (i.e., the elongation was towards the rotational axis). In this 
case, the beam half opening angle was given by

\begin{equation}
\label{EqCorrectionFactorParaFid} \rho_{\mathrm{ell}} = \frac{\rho}{\sqrt{(1 - \epsilon^2)\cos^2(\psi_{\mathrm{W}}) + \sin^2(\psi_{\mathrm{W}})}}    .
\end{equation}
In the second variant the major axis of the beam lay perpendicular 
to the fiducial plane (i.e., the elongation was in the direction of rotation) and the beam half opening angle was given by 

\begin{equation}
\label{EqCorrectionFactorPerpFid} \rho_{\mathrm{ell}} = \frac{\rho}{\sqrt{(1 - \epsilon^2)\sin^2(\psi_{\mathrm{W}}) + \cos^2(\psi_{\mathrm{W}})}}    .
\end{equation}
In both these equations $\rho$ is given by Eq.~\ref{EqBeamHalfOpeningAngle}, $\epsilon$ is the ellipticity of the beam and 
$\psi_{\mathrm{W}}$ is the magnetic longitude at which the line of sight enters 
and exits the beam, given by 

\begin{equation}
\label{EqPsiW} \sin(\psi_{\mathrm{W}}) = \frac{\sin(\alpha + \beta)\sin(W_{\mathrm{open}}/2)}{\sin(\rho_{\mathrm{ell}})} 
\end{equation}
(e.g., the appendix in \citealt{ww09}). 

Elliptical beams are predicted for inclined dipolar fields due to an effect known as meridional compression \citep{big90, mck93}. However, in that case $\rho$ forms the semi-major axis of the ellipse 
and the beam is compressed in the direction of the rotation axis. Correcting for such an effect will decrease the derived $\alpha$ values and hence was not considered.

\begin{figure}
\centering
\includegraphics[height=\hsize,angle=270]{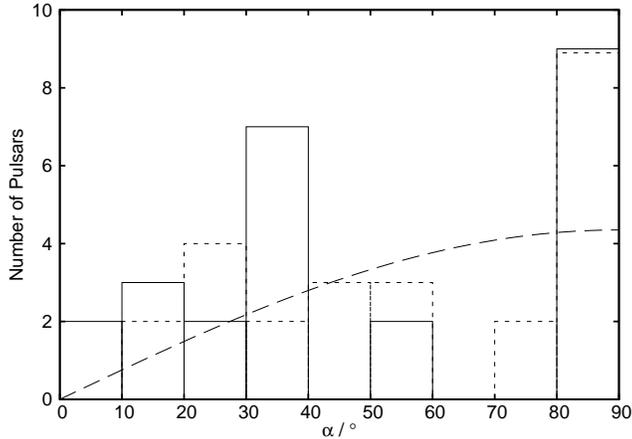}
\caption{\label{FigEllipticalBeamDistributions} Distribution of $\alpha$ values obtained assuming an elliptical beam with ellipticity $\epsilon$ = 0.94 elongated towards the rotational axis (solid line) and with $\epsilon$ = 0.86 elongated in the direction of rotation (dotted line), compared with a sinusoidal distribution (dashed curve). }
\end{figure}  

For each variant of the model a series of $\alpha$ distributions was 
determined corresponding to different $\epsilon$ values. As expected the effect 
of the elliptical beam was more pronounced for higher ellipticities. 
Fig.~\ref{FigEllipticalBeamDistributions} shows the $\alpha$ distributions corresponding to the respective best matches with a sinusoidal distribution.
These are found to be $\epsilon$ = 0.94 with meridional elongation and $\epsilon$ = 0.86 with longitudinal elongation, corresponding to axial ratios of 3 and 2 respectively. It can be seen that the distribution tends towards higher $\alpha$ values when the beam is elongated in the direction of rotation even though $\epsilon$ was lower in this case, as the line of sight will spend a longer time within the beam. However, both 
distributions still show marked departures from the sinusoidal 
distribution and the highest result returned by the KS test was $< 1\%$ for each variant, indicating a $\sim 3\sigma$ variation. Also, for many of the sample the enlarged beamwidth resulted in a constraint from the A/R effect which was inconsistent with the constraint from fitting of the RVM (in such cases $\alpha$ was assumed to be $90\degree$ in line with the methodology discussed in $\S$~4 of RWJ14). It is therefore apparent that the elliptical beam model cannot make the observed $\alpha$ distribution consistent with a sinusoidal distribution.

\subsection{Extra-cap emission}
\label{SectRandoms}

The final considered way to make the $\alpha$ distribution consistent with the sinusoidal distribution was ``extra-cap'' emission. In this situation the emission region 
is assumed to be circular, with a radius which exceeds the conventional polar 
cap radius by a factor $s$. This is similar to the elliptical beam 
model in that the beam half opening angle from 
Eq.~\ref{EqBeamHalfOpeningAngle} is an underestimate which leads 
to an underestimation of $\alpha$. Claims of extra-cap emission exist in the literature \citep{ww09, kjw+10}. In contrast, however, for PSR J0908--4913 \cite{kj08} found $s \leq 1$. It therefore seems that $s$ must differ between pulsars. Nevertheless, to compensate for the abundance of observed low $\alpha$ values in this scenario requires $s > 1$ for the majority of the sample.

To test this possibility the beam half-opening angles were corrected by applying the transformation 
$\theta_{\mathrm{PC}} \rightarrow s\theta_{\mathrm{PC}}$ to Eq.~\ref{EqBeamHalfOpeningAngle}. 
New $\alpha$ distributions were then determined using the same method 
detailed in $\S$~\ref{SectIntroduction}.

\begin{figure}
\centering
\includegraphics[height=\hsize,angle=270]{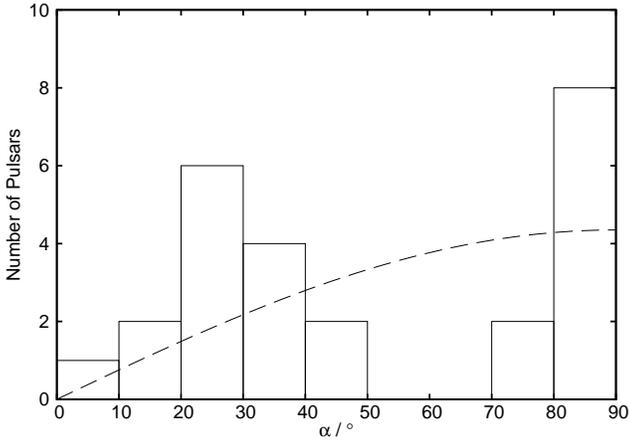}
\caption{\label{Figs200percent} Distribution of $\alpha$ values obtained by applying $s = 1.18$ to each pulsar. The sinusoidal distribution is also shown (dashed curve). }
\end{figure}

Firstly $\alpha$ distributions were obtained by applying the same $s$ value 
to each pulsar. Fig.~\ref{Figs200percent} shows the distribution 
corresponding to $s = 1.18$, which best fitted a sinusoidal distribution. The distributions are inconsistent, almost to the 3$\sigma$ level (a KS test returns a result of 0.94\%). The deficit of pulsars with $50\degree < \alpha < 70\degree$ cannot be compensated for in this scenario and also the optimum $s$ value is relatively close to $s = 1$, indicating that the effect on the original $\alpha$ distribution is marginal. At lower $s$, the low $\alpha$ peak shifted to lower values. When larger $s$ was used the number of pulsars for which the larger beam size resulted in a constraint from the A/R effect which was inconsistent with the constraint from RVM fitting increased. As these were assigned to $\alpha = 90\degree$ (see $\S$~4 of RWJ14), this led to larger discrepancies between the two distributions. 

Given the variation of $s$ values reported in the literature, various 
$s$ distributions 
were tested. For each test, $s$ values were drawn randomly from the 
distribution and applied to the observations. The peak of the corresponding 
$\alpha$ distributions occurred at higher $\alpha$ when the mean 
$s$ was higher and the $s$ distribution was narrower. However, even 
when a distribution with an unfeasibly high mean $s = 10$ was used, 
the resulting $\alpha$ distribution retained a peak at 
$\alpha \leq 60\degree$. This means that the $\alpha$ distribution 
cannot be made to be consistent with the sinusoidal distribution by 
choosing $s$ randomly from any distribution.

Also, consideration should be given to the $\alpha$ distributions 
determined by \cite{ran90} and \cite{gou94}, which were 
analysed by \cite{tm98}. These $\alpha$ values were calculated using 
empirically-determined relations between the component width and period 
with an assumed $\alpha$ dependence consistent with Eq.~\ref{EqRhoContours}, 
which did not rely on A/R effects or involve any direct assumption 
of $s$. The derived empirical relations are therefore independent 
of $s$, provided $s$ takes a constant value unrelated to $\alpha$. 
Given this, the existence of a similar skew in both Rankin's and Gould's 
$\alpha$ distributions provides more evidence that a simple scaling 
of $\rho$ via a parameter $s$ is not the full story. However, if $s$ 
is in some way related to the magnetic inclination, the $\alpha$-dependence 
of the component width will differ from that which was assumed by 
Rankin and by Gould. In this case their $\alpha$ values will have 
been similarly affected to the $\alpha$ values determined in RWJ14, 
potentially explaining the fact that the skew is apparent in all three 
samples. Therefore, it appears that if the observed excess of low 
$\alpha$ values is caused by the assumed $s$ values, then an 
$\alpha$-dependence of $s$ is required.

\subsection{Dependence of extra-cap emission on $\alpha$}
\label{SectSOfAlpha}

\begin{figure}
\centering
\includegraphics[height=\hsize,angle=270]{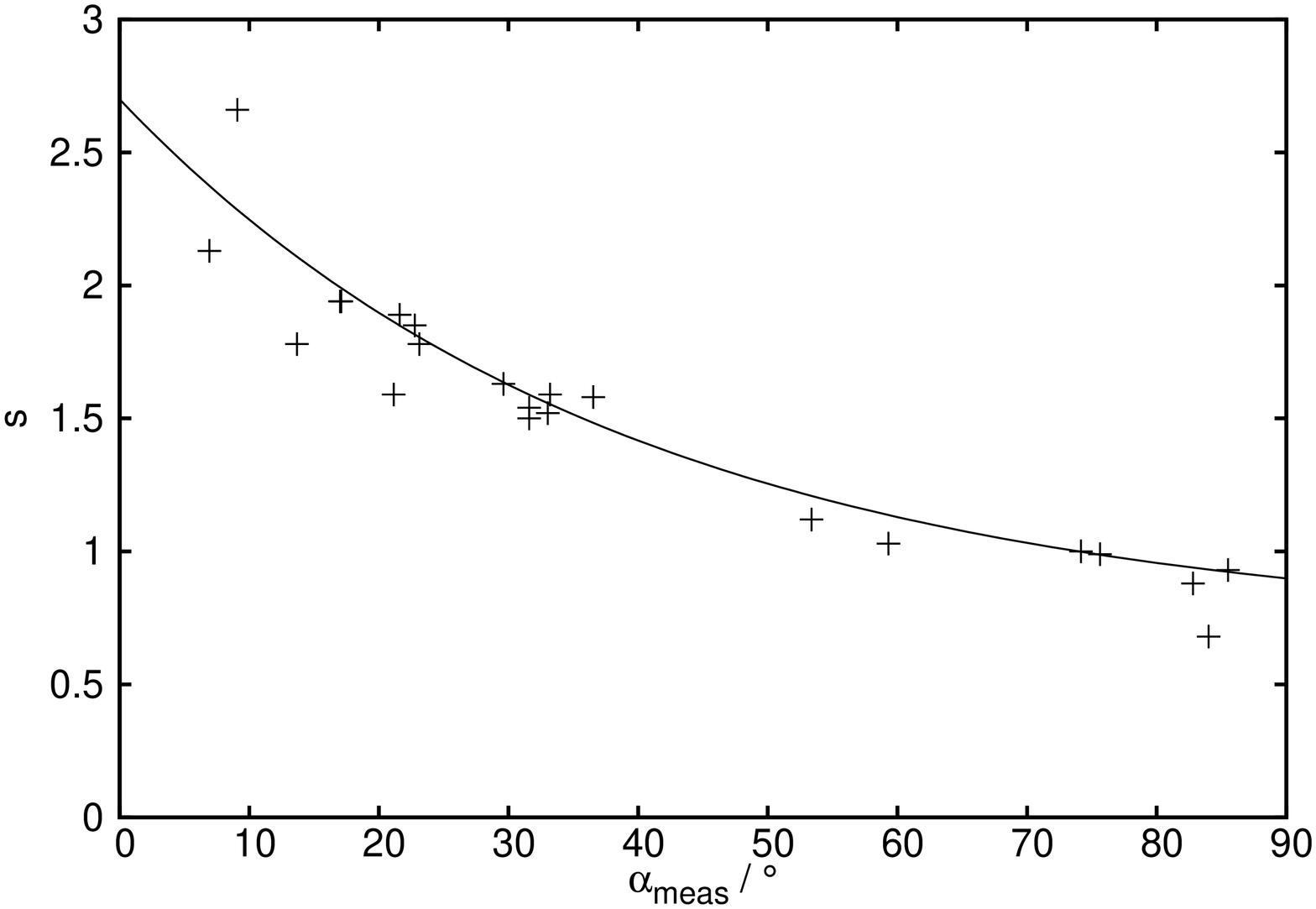}
\includegraphics[height=\hsize,angle=270]{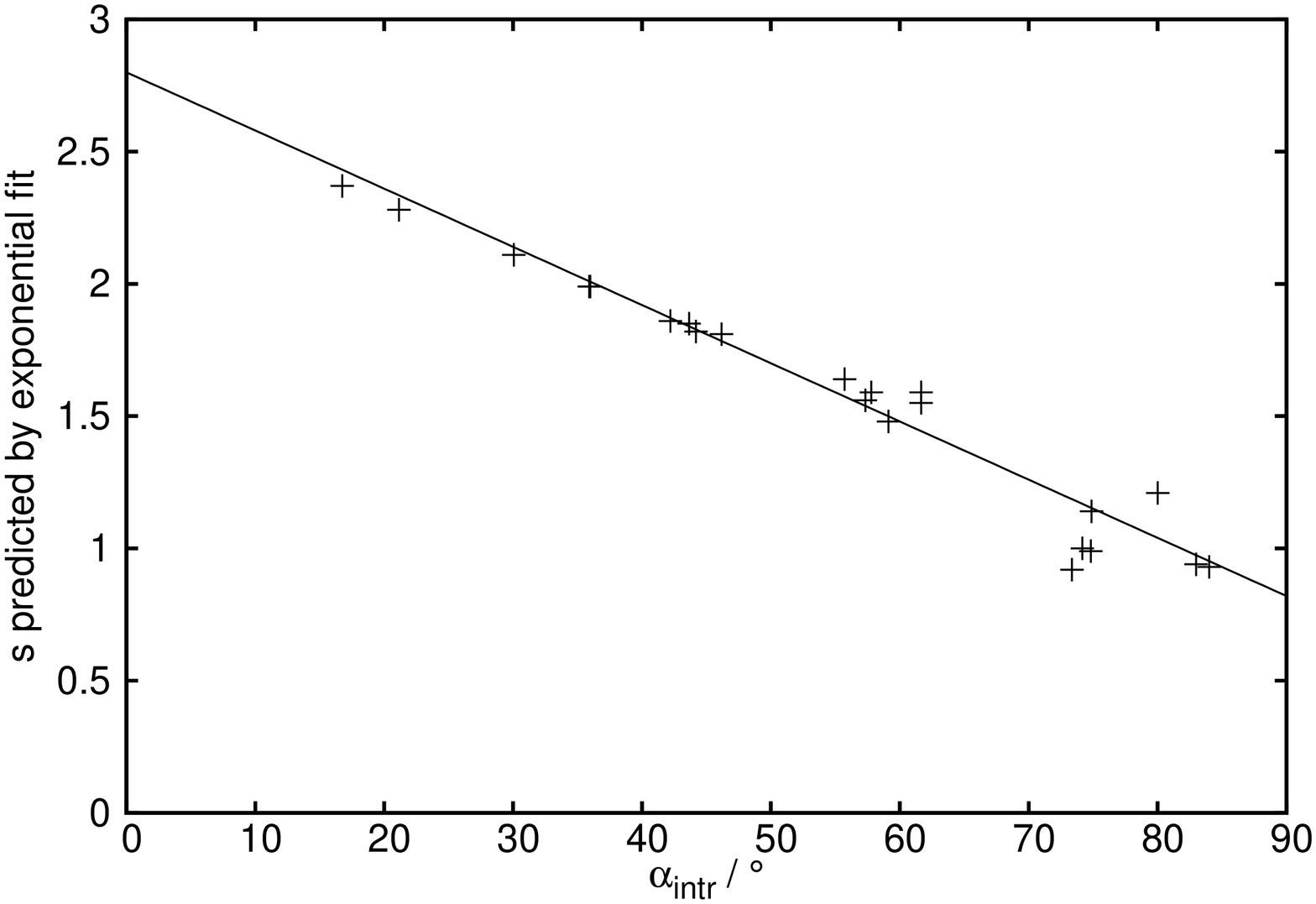}
\includegraphics[height=\hsize,angle=270]{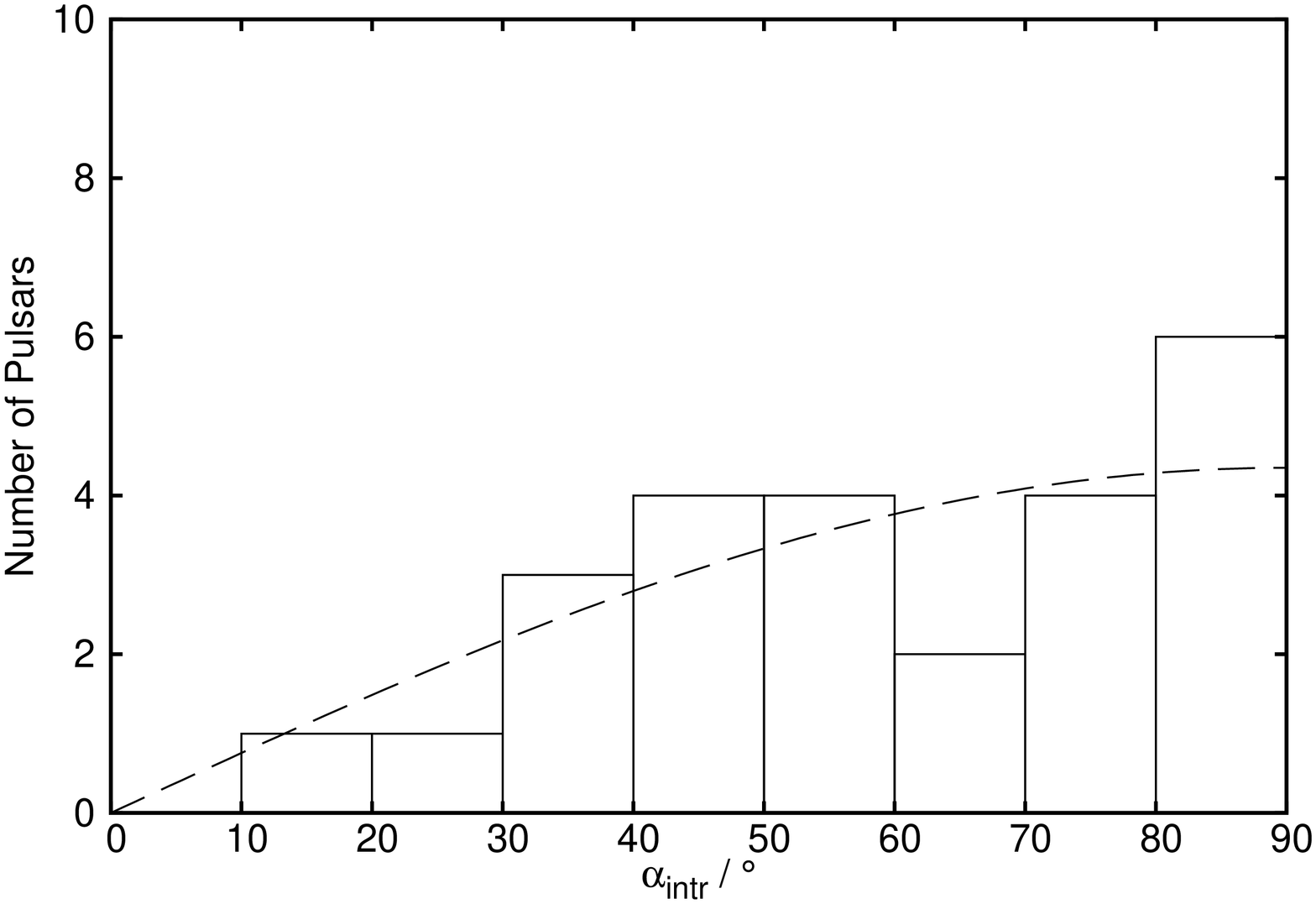}
\caption{\label{FigsOfAlpha} (Upper panel) The points indicate the value of $s$, the ratio of the emission region radius to the radius of the open-field-line region of a dipolar field, required for each pulsar such that the $\alpha$ distribution becomes sinusoidal, as a function of $\alpha_{\mathrm{meas}}$, the $\alpha$ value derived assuming $s = 1$. The data have been fitted with an exponential function. (Middle panel) The $s$ values predicted by the best fit shown in the upper panel, plotted as a function of $\alpha$ obtained after applying these $s$ values to the respective pulsars. This is therefore $s$ as a function of the intrinsic $\alpha$ under the assumption that the $\alpha$ distribution is sinusoidal. These data are well fitted by a linear function. (Lower panel) The distribution of these inferred intrinsic $\alpha$ values, which is indeed consistent with a sinusoidal distribution. }
\end{figure}

By choosing an optimum value of $s$ for each pulsar individually it is, by definition, possible to reproduce any desired $\alpha$ distribution. However, in $\S$~\ref{SectRandoms} we concluded that, if it is the cause of 
a systematic underestimation of $\alpha$, $s$ cannot be drawn randomly from a distribution and hence must either directly or indirectly depend on $\alpha$. To 
investigate what $\alpha$ dependence of $s$ would be able to reproduce a 
sinusoidal distribution, we made the assumption that the $\alpha$ 
values derived in RWJ14 are in the correct order. In other words, 
the pulsar with the lowest derived $\alpha$ was assumed to have the lowest 
intrinsic $\alpha$ of the sample and that there should be a monotonic relation between the intrinsic and measured $\alpha$. Keeping the pulsars in the same $\alpha$ order, an $s$ value was assigned to each pulsar such that the sinusoidal distribution was recreated. The upper panel of Fig.~\ref{FigsOfAlpha} shows the $s$ values required for this as a function of $\alpha_{\mathrm{meas}}$, the measured 
$\alpha$ values as derived under the assumption that $s = 1$.  

The trend in the upper panel of Fig.~\ref{FigsOfAlpha} is well fitted 
by an exponential function. The best exponential fit, which is also 
shown in the figure, is 

\begin{equation}
\label{EqSOfMeasuredAlpha} s(\alpha_{\mathrm{meas}}) = 2.0 e^{-\alpha_{\mathrm{meas}}/39\degree} + 0.7    .
\end{equation}
The errors on the fit parameters have not been quoted as the main source of error in this relation is likely to be unknown systematics. As expected, pulsars with lower $\alpha_{\mathrm{meas}}$ values have, 
on average, larger $s$ values in order to reduce the excess of low 
$\alpha$ values. There are several possible reasons for the scatter 
of data points around this fit, such as uncertainties in the determination 
of $\alpha_{\mathrm{meas}}$. Another possibility is that $s$ is not 
solely dependent on $\alpha$ but is also dependent on other parameters, 
thereby adding a random element. 

It is possible to use this relationship between $\alpha_{\mathrm{meas}}$ 
and $s$ to determine the intrinsic $\alpha$ for each pulsar. 
The $\alpha_{\mathrm{meas}}$ value of each pulsar and Eq.~\ref{EqSOfMeasuredAlpha} were used to predict $s$. This $s$ value was then applied 
to the pulsar to determine a ``corrected'' $\alpha$ value, $\alpha_{\mathrm{intr}}$, which should represent the true value of $\alpha$. The middle panel in Fig.~\ref{FigsOfAlpha} shows $s$ plotted as a function of $\alpha_{\mathrm{intr}}$, and the best linear fit to the data,

\begin{equation}
\label{EqSOfCorrectedAlpha} s(\alpha_{\mathrm{intr}}) = -0.022 \alpha_{\mathrm{intr}} + 2.80    ,
\end{equation} 
where the errors have been discarded as in Eq.~\ref{EqSOfMeasuredAlpha}. As noted above, any $\alpha$ distribution could be produced given the relevant set of $s$ values. However, it is encouraging (and non-trivial) that the $s$ values required to produce the sinusoidal distribution are a simple linear function of the pulsars' intrinsic $\alpha$. In addition the half-opening angle of the beam is approximately that expected from the last open field lines of a dipolar field (s = 1) for orthogonal rotators. 

The bottom panel in Fig.~\ref{FigsOfAlpha} contains the distribution of $\alpha_{\mathrm{intr}}$ values, which shows a good match with the sinusoidal distribution. The match is confirmed by a KS test between these values and the sinusoidal distribution, which returns 66\%. This is the highest KS test result of any scenario considered in this paper.

\subsection{Extra-cap emission - comparison with individual pulsars}
\label{SectIndividualPulsars}

\begin{figure}
\centering
\includegraphics[height=\hsize,angle=270]{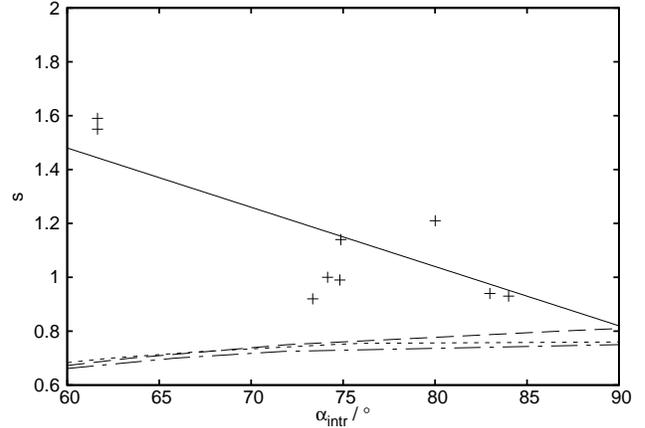}
\caption{\label{FigAlphaTracks} The three curves show the relationship between $s$ and $\alpha$ for PSRs J0729--1448 (dashed curve), J0742--2822 (dotted curve) and J1105--6107 (dash-dotted curve). Data points from the middle panel of Fig.~\ref{FigsOfAlpha} have been included to indicate the magnitude of the scatter about the best linear fit (solid line). }
\end{figure}

In RWJ14 three cases\footnote{PSRs J0729--1448, J0742--2822 and J1105--6107.} were found for which Eqs.~\ref{EqSOfMeasuredAlpha} and \ref{EqSOfCorrectedAlpha} cannot be used as no consistent solutions were found under the 
assumption that $s = 1$ and hence $\alpha_{\mathrm{meas}}$ could not be defined. This could be due to slight errors in the chosen fiducial plane or pulse edge position, or alternatively could be explained by a value $s < 1$. By calculating $\alpha_{\mathrm{intr}}$ using the method outlined in $\S$~\ref{SectTheory} as $s$ is varied, it is possible to plot a track for a given pulsar in ($\alpha_{\mathrm{intr}}$, $s$) space. The plotted track will only be consistent with 
the linear fit for a limited range of $\alpha$. Fig.~\ref{FigAlphaTracks} shows the tracks for these three pulsars, along with the best linear fit of $s(\alpha_{\mathrm{intr}})$ and some of the data points from the middle panel of Fig.~\ref{FigsOfAlpha}. It can be seen that for these pulsars the tracks are consistent with the linear fit (within the scatter of data points) for $\alpha \gtrsim 80\degree$. This suggests these are close to being orthogonal rotators, consistent with the conclusions of RWJ14.

\begin{figure}
\centering
\includegraphics[height=\hsize,angle=270]{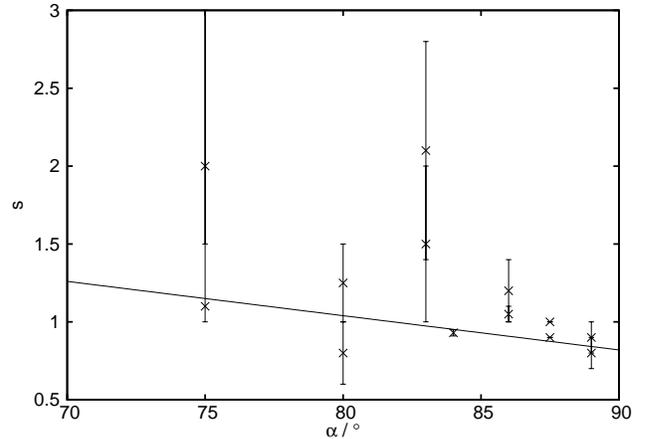}
\caption{\label{FigBonusInterpulsePulsars} Values of $\alpha$ and $s$ taken from the literature for PSR J0908--4913 (at $\alpha = 84\degree$), PSR J1057--5226 from \protect \cite{ww09} (at $\alpha = 75\degree$) and five pulsars analysed by \protect \cite{kjw+10}. Main pulse and interpulse values are shown for each pulsar. The solid line is the same linear fit as shown in the middle panel of Fig.~\ref{FigsOfAlpha}.}
\end{figure}

A useful test of the relationship between $s$ and $\alpha$ comes 
from interpulse pulsars. The presence of both a main pulse and interpulse 
often allows a significant $s$-independent constraint on $\alpha$ 
from RVM fitting alone. Combining this with an estimate 
of $s$ allows these pulsars to be placed directly onto a plot of 
$s$ versus $\alpha$ and their consistency with Eq.~\ref{EqSOfCorrectedAlpha}
can be checked. Values of $s$ and $\alpha$ for five interpulse 
pulsars which are not in our sample can be found in \cite{kjw+10}. 
These, along with PSRs J0908--4913 \citep{kj08} and the results for 
the two poles of PSR J1057--5226 \citep{ww09} are shown in 
Fig.~\ref{FigBonusInterpulsePulsars}. In the latter paper $s$ was 
found for each pole as a function of the emission height. The behaviour 
of those functions allows the lower limits on $s$ to be estimated,
but there appears to be no upper limit for either pole. 

It can be seen that these points are mostly consistent with the trend shown in Fig.~\ref{FigsOfAlpha}, although with some scatter. For most of the pulsars in this figure, the value of $s$ appears to differ between the main pulse and interpulse. This is consistent with the scatter of data points around the fit seen in the upper and middle panels of Fig.~\ref{FigsOfAlpha} and may be evidence for an additional dependence of $s$ on another parameter, or a random element. It should be noted that as these pulsars only cover a small range of $\alpha$ the gradient of the relationship cannot be confirmed. A greater test would be possible if independent $s$ and $\alpha$ values were also known for more closely aligned pulsars.

\section{Discussion}
\label{SectDiscussion}

In the previous sections we have shown that the distribution of $\alpha$ values presented in \cite{rwj14a} (RWJ14) differs significantly from the sinusoidal distribution one may expect for this sample of young, $\gamma$-ray-loud pulsars. We have also considered various biases which might be the cause of this difference. In this section we will first consider, in a model-independent way, the implications of a scenario in which the distribution is intrinsically sinusoidal for this sample. We devise a simple scheme to infer the intrinsic $\alpha$ value from observations which are affected by a bias which affects the measurements. We will then discuss the plausibility and implications of the two possible biases, a systematic misplacement of the fiducial plane and an $\alpha$-dependence of $s$, which have been found to be capable of reproducing an approximately sinusoidal distribution from the measured distribution. Finally we will consider the alternative scenario, in which the intrinsic $\alpha$ distribution is skewed to low $\alpha$ values compared to a sinusoidal distribution.

\subsection{A model-independent relation between measured and intrinsic $\alpha$ in the case of random axis orientation}
\label{SectAlphaAlpha} 

\begin{figure}
\centering
\includegraphics[height=\hsize,angle=270]{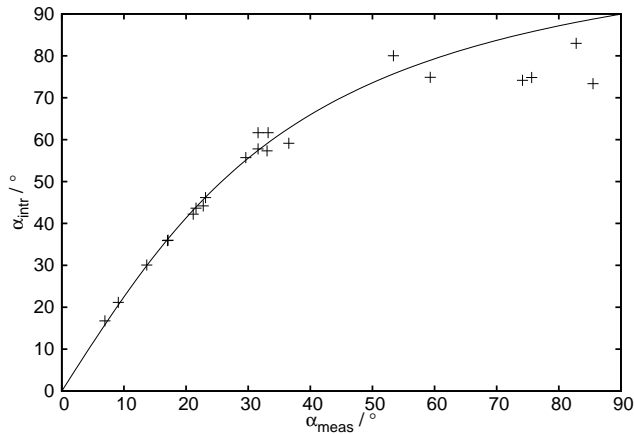}
\caption{\label{FigAlphaOfAlpha} The inferred intrinsic magnetic inclination, $\alpha_{\mathrm{intr}}$, as a function of the measured magnetic inclination, $\alpha_{\mathrm{meas}}$. These inferred values are those required to reproduce a sinusoidal distribution, assuming a monotonic relation with the measured inclination. An arctangential fit to the data is also shown. }
\end{figure}

If the rotation and magnetic axes are randomly orientated, and hence the $\alpha$ distribution is intrinsically sinusoidal, it is desirable to find a relation between the measured and intrinsic $\alpha$ values. Fig.~\ref{FigAlphaOfAlpha} shows $\alpha_{\mathrm{intr}}$, as determined in $\S$~\ref{SectSOfAlpha} under the assumption that $s$ follows Eq.~\ref{EqSOfCorrectedAlpha}, as a function of $\alpha_{\mathrm{meas}}$, as determined in RWJ14 under the assumption 
$s = 1$. However, it should be noted that any model used to 
recreate a sinusoidal distribution from the observed $\alpha$ 
values would give a similar diagram.

These data are relatively well fitted by an arctangential function. At high $\alpha$ the $\alpha_{\mathrm{meas}}$ values are higher than the $\alpha_{\mathrm{intr}}$ values. This is because $s$ is slightly smaller than 1 for orthogonal rotators according to Eq.~\ref{EqSOfCorrectedAlpha}. However, it follows from Eq.~\ref{EqRhoContours} that $\alpha_{\mathrm{intr}}$ is more sensitive to $s$ (and, via Eq.~\ref{EqSOfMeasuredAlpha}, to errors on $\alpha_{\mathrm{meas}}$) at values close to $90\degree$. Therefore, accounting for the subsequent uncertainty in $\alpha_{\mathrm{intr}}$, the data in the figure are consistent with a relation which passes through the point $\alpha_{\mathrm{intr}} = \alpha_{\mathrm{meas}} = 90\degree$. This is desirable in order to find a relation which allows for the prediction of orthogonal rotators. This was taken as a boundary condition for the fit. The best arctangential fit (shown in Fig.~\ref{FigAlphaOfAlpha}) was found to be

\begin{equation}
\label{EqAlphaOfAlpha} \alpha_{\mathrm{intr}} = 1.27\degree \arctan\left(\frac{\alpha_{\mathrm{meas}}}{31.25\degree}\right)    .
\end{equation} 
The scatter of data points demonstrates that there is significant uncertainty when applying this conversion to any particular pulsar. However, this equation can nevertheless be used as a population-averaged conversion between the measured and intrinsic $\alpha$ values. It should be noted that, although the scatter of data points can be used to estimate errors on the fit parameters, the total error is likely to be dominated by unknown systematics related to the determination of $\alpha_{\mathrm{meas}}$. For this reason we have chosen to disregard the errors from Eq.~\ref{EqAlphaOfAlpha}.

The relation is effectively independent of the method used to determine $\alpha_{\mathrm{intr}}$, as any correction to the measured $\alpha$ which results in a sinusoidal distribution will yield a similar set of values. The only assumption (other than that the intrinsic distribution is sinusoidal) to which the relation is sensitive is that higher measured $\alpha$ values in general correspond to higher intrinsic $\alpha$ values, or in other words that the PA curve contains at least some information about $\alpha$.

\begin{table*}
\footnotesize
\centering
\setlength{\extrarowheight}{1 mm}
\caption{\label{TabCorrectedViewingGeometries} The allowed and favoured viewing geometries for the sample after correcting the values (given in Table~2 of RWJ14) of $\alpha$ according to Eq.~\ref{EqAlphaOfAlpha} and $\beta$ according to Eq.~\ref{EqBetaCorrection} (i.e., assuming the magnetic and rotation axes to be randomly orientated). MP and IP refer to $\alpha$ and $\beta$ values with respect to the main pulse and interpulse. In the case of PSR J1119--6127, the two cases (a) and (b) are described in RWJ14. It should be noted that values of $\alpha > 90\degree$ have not been mapped into $0 < \alpha < 90\degree$. }

\begin{tabular}{l>{\hspace{6 mm}}r@{ - }lr@{ - }lrr}
\hline
PSR & \multicolumn{4}{c}{Allowed Solutions} & \multicolumn{2}{c}{Favoured Solutions} \\
& \multicolumn{2}{c}{$\alpha_{\mathrm{intr}}$ / $\degree$} & \multicolumn{2}{c}{$\beta_{\mathrm{intr}}$ / $\degree$} & $\alpha_{\mathrm{intr}}$ / $\degree$ & $\beta_{\mathrm{intr}}$ / $\degree$ \\
\hline
J0631+1036 & 59 & 127 & --10.5 & --4 & 92\hspace{13 pt} & --10\hspace{13 pt} \\
J0659+1414 & 58 & 139 & --22 & --8 & 101\hspace{13 pt} & --19\hspace{13 pt} \\
J0729--1448 & 56 & 122 & 3 & 7 & 90\hspace{13 pt} & 6\hspace{13 pt} \\
J0742--2822 & 55 & 180 & --7 & 0 & 90\hspace{13 pt} & --6.5\hspace{7 pt} \\
J0835--4510 & 66 & 98 & --7.5 & --7 & 85\hspace{13 pt} & --7.5\hspace{7 pt} \\
J0908--4913 (MP) & 96 & 96.8 & --8.5 & --6.3 & 96.1\hspace{7 pt} & --5.9\hspace{7 pt} \\
J0908--4913 (IP) & \multicolumn{2}{c}{} & \multicolumn{2}{c}{} & 83.9\hspace{7 pt} & 6.3\hspace{7 pt} \\
J0940--5428 & \multicolumn{2}{c}{0 - 73; 102 - 180} & 0 & 21 & 117\hspace{13 pt} & 18\hspace{13 pt} \\
J1016--5857 & \multicolumn{2}{c}{0 - 65; 119 - 180} & --13 & 0 & 144\hspace{13 pt} & --6\hspace{13 pt} \\
J1048--5832 & 0 & 74 & 0 & 9 & 55\hspace{13 pt} & 8\hspace{13 pt} \\
J1057--5226 (MP) & 68 & 92 & 10 & 48 & 86\hspace{13 pt} & 20\hspace{13 pt} \\
J1057--5226 (IP) & \multicolumn{2}{c}{} & \multicolumn{2}{c}{} & 94\hspace{13 pt} & 12\hspace{13 pt} \\
J1105--6107 & 53 & 124 & 3 & 5 & 90\hspace{13 pt} & 4\hspace{13 pt} \\
J1112--6103 & 0 & 180 & --5.5 & 0 & 104\hspace{13 pt} & --5\hspace{13 pt} \\
J1119--6127 (a) & \multicolumn{2}{c}{0 - 80; 108 - 180} & --25 & 0 & 159\hspace{13 pt} & --9\hspace{13 pt} \\
J1119--6127 (b) & \multicolumn{2}{c}{0 - 72; 118 - 180} & --25 & 0 & 164\hspace{13 pt} & --7\hspace{13 pt} \\
J1357--6429 & \multicolumn{2}{c}{0 - 77; 93 - 180} & 0 & 55 & 16\hspace{13 pt} & 11\hspace{13 pt} \\
J1410--6132 & 0 & 180 & 0 & 5.5 & 121\hspace{13 pt} & 4.5\hspace{7 pt} \\
J1420--6048 & 0 & 59 & 0 & 13 & 36\hspace{13 pt} & 8\hspace{13 pt} \\
J1513--5908 & 0 & 180 & 0 & 90 & 30\hspace{13 pt} & 34\hspace{13 pt} \\
J1531--5610 & 0 & 180 & --59 & 0 & 136\hspace{13 pt} & --28\hspace{13 pt} \\
J1648--4611 & 0 & 180 & --14 & 0 & 137\hspace{13 pt} & --8\hspace{13 pt} \\
J1702--4128 & \multicolumn{2}{c}{0 - 83; 101 - 180} & --14 & 0 & 46\hspace{13 pt} & --10\hspace{13 pt} \\
J1709--4429 & 27 & 72 & 12 & 24 & 57\hspace{13 pt} & 21\hspace{13 pt} \\
J1718--3825 & \multicolumn{2}{c}{0 - 81; 97 - 148} & 0 & 17 & 46\hspace{13 pt} & 10\hspace{13 pt} \\
J1730--3350 & 0 & 180 & --10 & 0 & 123\hspace{13 pt} & --6\hspace{13 pt} \\
J1801--2451 & \multicolumn{2}{c}{0 - 78; 107 - 180} & --14 & 0 & 121\hspace{13 pt} & --11\hspace{13 pt} \\
J1835--1106 & 0 & 180 & 0 & 13 & 89\hspace{13 pt} & 11\hspace{13 pt} \\
\hline

\end{tabular}
\end{table*}

It should be noted that Eq.~\ref{EqAlphaOfAlpha} is derived assuming an intrinsically sinusoidal $\alpha$ distribution. As discussed in $\S$~\ref{SectIntroduction}, it is expected that due to the selection effects of increased $\gamma$-ray intensity modulation and radio beaming fraction when $\alpha$ is larger, this sample should contain more high-$\alpha$ values compared to a sinusoidal distribution. In this case, the relation between $\alpha_{\mathrm{intr}}$ and $\alpha_{\mathrm{meas}}$ would rise more steeply at low measured $\alpha$. Eq.~\ref{EqAlphaOfAlpha} may still be used as a first order correction and the mentioned biases will make the intrinsic $\alpha$ values higher, making the correction presented here somewhat conservative.

A change in $\alpha$ will be accompanied by a change in $\beta$ in order to make the observed PA curve consistent with the RVM. However, it is possible, once $\alpha$ has been corrected, to correct $\beta$ accordingly. The two angles can be related to the maximum gradient of the measured PA curve via the equation

\begin{equation}
\label{EqdPAdPhi} \left(\frac{d\psi}{d\phi}\right)_{\mathrm{MAX}}=\frac{\sin\alpha}{\sin\beta}
\end{equation} 
\citep{kom70}. Use of this relation is justified as the gradient is, in most cases, the most constrained aspect of the RVM. The right-hand side should be identical in the cases of the measured and intrinsic viewing geometries, leading to

\begin{equation}
\label{EqBetaCorrection} \sin\beta_{\mathrm{intr}}=\frac{\sin\alpha_{\mathrm{intr}}}{\sin\alpha_{\mathrm{meas}}}\sin\beta_{\mathrm{meas}}    .
\end{equation} 
As an example of this method of correction, Table~\ref{TabCorrectedViewingGeometries} contains the viewing geometry constraints and favoured $\alpha$ and $\beta$ values corrected using Eqs.~\ref{EqAlphaOfAlpha} and \ref{EqBetaCorrection} from those given in Table 2 of RWJ14. Note that in all cases $\beta_{\mathrm{intr}}$ is larger than $\beta_{\mathrm{meas}}$.

\subsection{A systematic fiducial plane offset}
\label{SectImplicationsPhiFidOffset}

In $\S$~\ref{SectMovingPhiFid} we found that a systematic offset of $\sim 14\degree$ in the fiducial plane position could explain the difference between the measured $\alpha$ distribution and a sinusoidal distribution. The fixed offset in pulse phase implies that even in cases with high mirror symmetry in the profile (such as PSRs~J1420--6048 and J1648--4611) the beam is not symmetrically illuminated.

One might expect that the offset of the fiducial plane is dependent on the pulse width. We found that the observed bias towards low values could be explained if only the trailing 45\% of each beam is illuminated. In the case of PSR~J1057--5226, \cite{ww09} argued that the main pulse represents only the trailing 50\% of the beam. This suggests that at least some pulsars should be capable of having the required asymmetry in the illumination.

A problem arises when accounting for the selection effects expected due to $\gamma$-ray detectability and the radio beaming fraction (see $\S$~\ref{SectIntroduction}). Both effects predict the intrinsic distribution to have an excess of large $\alpha$ values relative to a sinusoidal distribution. However, neither a constant offset nor an offset as a fixed proportion of the pulse width could reproduce a distribution with such an excess.

Another problem with a systematic misplacement of the fiducial plane comes from the literature. A systematic error in the fiducial plane position would affect our results as this position was used to calculate $\rho$ and $W_{\mathrm{open}}$. However, a similar bias was found in the $\alpha$ distributions of \cite{ran90} and \cite{gou94} (as shown by \citealt{tm98}). As explained in $\S$~\ref{SectRandoms}, both authors used relations which are not affected by the choice of the fiducial plane position, and so the $\alpha$ values they derived should be independent of a systematic effect on the location of the fiducial plane. In light of this, the presence of a similar bias in the distributions as shown in \cite{tm98} strongly suggests that the effectiveness of both methods of offsetting the fiducial plane in retrieving a sinusoidal distribution from our data is coincidental, and cannot explain the departure of the observed $\alpha$ distribution from a sinusoidal distribution.

\subsection{Implications of an $\alpha$-dependence of $s$}

An $s$ value which is dependent on $\alpha$ is a more plausible explanation for the difference between the measured and sinusoidal distributions than that discussed in the previous subsection. The additional $\alpha$ dependence which this situation would introduce would affect the $\alpha$ values as measured in RWJ14, as well as those derived by, for instance, \cite{ran90} and \cite{gou94}. This could therefore explain the presence of a low-$\alpha$ bias in all three samples.

Extra-cap emission has been suggested in the literature and appears to be unavoidable in some cases (e.g., the main pulse of PSR~J1057--5226). Values of $s > 1$ mean either that emission can be generated on closed field lines or that some field lines traditionally thought to be closed are in fact open. The former has implications for the emission mechanism. For example, the \cite{rs75} model of the emission mechanism, which draws on the \cite{stu71} model for particle acceleration within the magnetosphere, requires the emitting particles to lie on open field lines; any emission generated on closed field lines would require a different mechanism. 

The latter situation has implications for the magnetic field structure. The open-field-line region we have assumed in this paper is based on a static dipolar field. We have not included higher order effects such as relativistic deformations of the field close to the light cylinder (e.g., \citealt{mic73}), which would alter the boundary of the open-field-line region. This boundary would be a natural location for the edge of the beam, making this interpretation more palatable. However, not every conceivable correction to the field structure would yield the linear dependence of $s$ on $\alpha$ suggested by our data ($\S$~\ref{SectSOfAlpha}). It should be noted that our results are insensitive to the shape of the beam, as our simulations of elliptical beams show (see $\S$~\ref{SectNonCirc}). If the beams are elliptical, $s$ in Eq.~\ref{EqSOfCorrectedAlpha} should be considered to correspond to the semi-minor axis of the beam.

Extra-cap emission also has implications for the compatibility of radio and $\gamma$-ray models. The $\gamma$-ray models with extended emission regions in the outer parts of the magnetosphere, such as the outer gap model \citep{crz00} or slot gap model \citep{mh04a}, typically require large values of $\alpha$ for pulsed emission to be observed (e.g. \citealt{wrw+09}). This is in stark contrast with the $\alpha$ values derived from radio polarisation measurements (Fig.~\ref{FigOriginalDistribution}), for which low values appear to be preferred. Making the radio beam wider will increase the radio-derived $\alpha$ values (c.f. Fig.~\ref{FigAlphaOfAlpha}), resulting in closer agreement with the expectation from $\gamma$-ray models. Note that Fig.~\ref{FigsOfAlpha} demonstrates that extra-cap emission is able to reproduce a sinusoidal $\alpha$ distribution. The fact that these pulsars are $\gamma$-ray-detected might suggest that the actual $\alpha$ distribution for this sample should be skewed to higher values, which would imply that the effect of extra-cap emission is stronger than implied by the equations in $\S$~\ref{SectSOfAlpha} and \ref{SectAlphaAlpha}. A second effect of extra-cap emission is that the $\beta$ values allowed by radio modeling increase. This increase in the parameter space available to the $\gamma$-ray models will increase the potential for models to match the shape of the observed light curves.

\subsection{Implications of an intrinsic low-$\alpha$ bias}

Equally interesting is the possibility that the measured $\alpha$ distribution (Fig.~\ref{FigOriginalDistribution}) would, after correcting for the selection effects discussed in $\S$~\ref{SectIntroduction}, accurately represent the birth distribution of $\alpha$ values for this sample of pulsars and hence that no additional corrections are needed. This would have implications for determined timescales for alignment of the magnetic axis. The method used by \cite{wj08a}, for example, assumes random orientation of the axes at the birth of the neutron star; if the axes in fact already tend towards alignment at birth a longer alignment timescale than their quoted value $7 \times 10^7$~yr is required. In contrast to this, the method used by \cite{tm98} (in which the average $\alpha$ value of a sample was analysed as a function of characteristic age) was not sensitive to the birth distribution. \cite{ycb+10} calculated alignment timescales for various models with differing birth $\alpha$ distributions and showed that their value of $10^6$~yr is fairly insensitive to the initial $\alpha$ distribution.

The sample analysed in this paper consists of young $\gamma$-ray-detected pulsars. If such a birth distribution is intrinsic \emph{only} to $\gamma$-ray-loud pulsars (i.e., the $\gamma$-ray-quiet population is still sinusoidal), this would be an important distinction and could allow insight into the differences between pulsars which emit $\gamma$-rays and those which do not. 

Alternatively, this birth distribution may apply to the pulsar population as a whole. This would imply the cause of the bias towards low $\alpha$ is the supernova (that is, some characteristic of the supernova favours closer alignment of the magnetic and rotation axes). The presence of a similar bias in the $\alpha$ distributions of \cite{tm98} suggest that this could be the case, as those pulsars were not selected according to detectability in $\gamma$-rays and so are a mixture of $\gamma$-ray-loud and -quiet sources.

\section{Conclusions}
\label{SectConclusions} 

\cite{rwj14a} presented the observed distribution of $\alpha$, 
the magnetic inclination, for a sample of young $\gamma$-ray-loud 
pulsars. This involved utilising several common assumptions 
relating to the structure and alignment of the magnetic 
field. This distribution is not consistent 
with the sinusoidal distribution expected if the 
magnetic and rotation axes of a neutron star are randomly orientated 
at birth, confirming the results of several previous studies. The observed distribution is skewed towards low values, which is opposite to the skew which would be expected from considerations of the radio beaming fraction or from the prediction by $\gamma$-ray models that pulsars are more easily detected in $\gamma$-rays when $\alpha$ is large.
Assuming that the intrinsic distribution is sinusoidal we have explored a number of potential causes, including the effects of systematic underestimation 
of the pulse phase of the inflection point of the PA curve, 
systematic overestimation 
of the fiducial plane position, a possible emission height gradient, 
a possible elliptical emission beam and various distributions of 
the ratio $s$ between the radius of a circular emission region and 
that expected for the open-field-line region of a dipolar field. As a result 
we are left with two alternative scenarios: either the birth $\alpha$ 
distribution is intrinsically non-sinusoidal such that there is a 
significant excess of low-$\alpha$ pulsars, or the measured $\alpha$ 
is not an accurate representation of the intrinsic values. 
Under the assumption of an intrinsically random orientation of the rotation and magnetic axes, the measured and intrinsic $\alpha$ and $\beta$ values may be related by Eqs.~\ref{EqAlphaOfAlpha} and \ref{EqBetaCorrection} respectively. 
However, as discussed, the intrinsic $\alpha$ distribution is likely 
to be skewed towards higher values relative to the sinusoidal distribution, 
in which case these relations will be a somewhat conservative
first order correction.   

We argue the discrepancy between the measured and intrinsic $\alpha$ 
values can be explained by a linear dependence of $s$ on the intrinsic 
$\alpha$ (Eq.~\ref{EqSOfCorrectedAlpha}). This relation predicts that the emission region is larger than expected for the standard assumption of emission confined to the open-field-line region of a dipolar field when $\alpha$ is small, tending towards $s \approx 1$ as the intrinsic $\alpha = 90\degree$. Although there must by definition be an $\alpha$-dependent $s$ distribution which will relate any desired $\alpha$ distribution to the observed distribution, the fact that our result is a simple linear relation is non-trivial. It is also non-trivial that the size of the emission region tends towards that expected for a dipolar field for orthogonal rotators.

Two alternative effects were considered which could bias the $\alpha$ 
distribution: that the fiducial plane position is systematically 
$\sim 14\degree$ earlier in rotational phase than our 
expectations based on the profile morphology, and that 
only the trailing 45\% of the beam is illuminated. These were both 
found to be mathematically possible but it has been argued to be implausible that either is the sole cause. We 
do not rule out the possibilities of elliptical beams or an emission 
height gradient, but find that these are not sufficient to explain 
the departure of the observed $\alpha$ distribution from a sinusoidal 
distribution.

Both an intrinsically non-sinusoidal birth $\alpha$ distribution and a discrepancy between the intrinsic and measured distributions have important implications for population 
studies. If the $\alpha$ distribution is intrinsically non-sinusoidal 
the population-averaged beaming fraction will be lower. 
In the case of an $\alpha$-dependent $s$ value the 
beaming fraction of each pulsar will be affected differently, but 
the population-averaged value will be made higher. 
There are also implications for the timescale for possible 
alignment of the two axes as a neutron star ages. Both scenarios 
suggest this timescale to be longer than some previous estimates.

Each scenario raises important questions about neutron 
star physics. An $\alpha$-dependence of $s$ suggests complexities 
in the radio emission process or magnetic field structure. An 
intrinsically non-sinusoidal birth $\alpha$ distribution has implications 
for the formation of the neutron star or for the 
dependence of the $\gamma$-ray emission process on magnetic 
inclination making \emph{aligned} rotators easier to detect.

Finally we note that an $\alpha$-dependent $s$ would 
make the radio models more compatible with the preferred $\gamma$-ray 
models, which place the extended emission regions far above the 
polar cap. This is because extra-cap emission will increase both 
the radio-derived $\alpha$ and $\beta$ values. Larger $\alpha$ 
values are favoured by the $\gamma$-ray models, while the possibility 
of larger $\beta$ values will increase the allowed parameter range 
available for the $\gamma$-ray models to fit the observed light 
curves. Extra-cap emission therefore should be seriously considered 
in radio beam models.

\section*{Acknowledgments}

We wish to thank the referee Aris Karastergiou for his constructive input.

~

\bibliographystyle{mn2e}

\bsp

\label{lastpage}

\end{document}